\documentclass[3p,times,twocolumn]{elsarticle}

\usepackage{ecrc}

\volume{00}

\firstpage{1}

\journalname{Nuclear Physics B Proceedings Supplement}

\runauth{}

\jid{nuphbp}

\jnltitlelogo{Nuclear Physics B Proceedings Supplement}

\usepackage{amssymb}

\usepackage[figuresright]{rotating}

\begin{document}

\begin{frontmatter}



\dochead{}

\title{Models and Phenomenology of Neutrino Masses circa 2010}


\author{Mu-Chun Chen}

\address{Department of Physics \& Astronomy, University of California, Irvine, CA 92697-4575, U.S.A.}

\author{K.T. Mahanthappa}

\address{Department of Physics, University of Colorado, Boulder, CO 80309-0390, U.S.A.}

\begin{abstract}
We review recent developments in models of neutrino masses and mixing. Emphases are given to models based on finite group family symmetries from which the tri-bimaximal neutrino mixing can arise. In particular, we describe one recent model based on SUSY SU(5) combined with a family symmetry based on the double tetrahedral group, $T^{\prime}$. All 22 observable fermion masses and mixing angles and CP violating measures are fitted with only 9 parameters. In this model, a near tri-bimaximal MNS matrix and a realistic CKM matrix are simultaneously generated;  the MNS matrix gets  slightly modified by virtue of having the Georgi-Jarlskog relations. Due to the presence of complex Clebsch-Gordan coefficients in $T^{\prime}$, CP violation in this model is entirely geometrical in origin.  The {\it prediction} of the model for the leptonic Dirac CP phase is $227^{o}$, which turns out to be very close to the current best fit value of $220^{o}$ from SuperK.
\end{abstract}


\end{frontmatter}



\section{Introduction}

One of the most significant unsolved questions in particle physics is the origin of fermion mass hierarchy and flavor mixing. Even though the Standard Model (SM) works beautifully in explaining all particle interactions, it has many free parameters in the Yukawa sector that accommodate the observed masses and mixing angles for quarks and leptons. The  number of free parameters can be greatly reduced by expanding the SM gauge group to a grand unified gauge symmetry, which gives rise to inter-family relations that connect quarks and leptons within the same family. Further reduction of parameters can be achieved with an additional family symmetry that relates quarks and leptons of different families. (For reviews, see {\it e.g.} \cite{Chen:2003zv}.)

The recent  advent of the neutrino oscillation data from Super-Kamiokande gives further support to models based on Grand Unified Theories (GUTs), in which the seesaw mechanism can arise  naturally. The global fit~\cite{GonzalezGarcia:2010er} to current data from neutrino oscillation experiments give the following best fit values and $1\sigma$ limits for the mixing parameters,
\begin{eqnarray}
\sin^{2} \theta_{12} \! \! & \! \! = \!\! & \!\! 0.319 \; (0.303 - 0.335) \; ,  \nonumber \\
\sin^{2} \theta_{23} \! \! & \!\! = \!\! & \!\!  0.463 \; (0.415 - 0.530) \; , \nonumber \\
\sin \theta_{13} \! \! & \!\! = \!\! & \!\! 0.127 \;  (0.072 - 0.165) \; , \nonumber 
\end{eqnarray}
while for the mass squared differences, the global fit gives,
\begin{eqnarray}
\Delta m_{12}^{2} \!\!& \!\!= \!\!&\!\! (7.59 \pm 0.20) \times 10^{-5} \; \mbox{eV}^2 \; , \nonumber\\
\Delta m_{31}^{2} \!\!& \!\! =\!\! &\!\! \bigg\{
\begin{array}{l}
(2.46 \pm 0.12)  \times 10^{-3} \; \mbox{eV}^2 \; , \\
-(2.36 \pm 0.11) \times 10^{-3} \; \mbox{eV}^{2} \; , 
\end{array} \nonumber
\end{eqnarray}
where the normal hierarchy corresponds to the global minima in the fit. 
These experimental best fit values for the mixing parameters are very close to the values arising from the so-called ``tri-bimaximal'' mixing (TBM) matrix~\cite{Harrison:2002er},
\begin{equation}
U_{\mathrm{TBM}} = \left(\begin{array}{ccc}
\sqrt{2/3} & 1/\sqrt{3} & 0\\
-\sqrt{1/6} & 1/\sqrt{3} & -1/\sqrt{2}\\
-\sqrt{1/6} & 1/\sqrt{3} & 1/\sqrt{2}
\end{array}\right) \; , \label{eq:tri-bi}
\end{equation}
which predicts 
\begin{eqnarray} \!\!\!\!\!
\sin^{2}\theta_{\mathrm{\scriptscriptstyle atm}}^{\mathrm{\scriptscriptstyle TBM}} =  1/2, \; 
\sin^{2}\theta_{\odot}^{\mathrm{\scriptscriptstyle TBM}}  =  1/3, \; 
\sin\theta_{13}^{\mathrm{\scriptscriptstyle TBM}} =  0.
 \end{eqnarray}
Even though the predicted $\theta_{\odot, \mathrm{TBM}}$ is currently still allowed by the experimental data at $2\sigma$, as it is very close to the upper bound at the $2\sigma$ limit, it  may be ruled out once more precise measurements are made in the upcoming experiments. In addition, the global has indicated a hint for non-zero $\theta_{13}$~\cite{Fogli:2008jx,GonzalezGarcia:2010er}.  Some small deviation from the exact tri-bimaximal mixing pattern is thus needed to agree with the experimental best fit values. As illustrated below in a model based on SUSY $SU(5) \times T^{\prime}$~\cite{Chen:2007afa,Chen:2009gf}, such deviation is predicted as a result of the SU(5) GUT relations. 
 
\section{Finite Group Family Symmetries}

The tri-bimaximal neutrino mixing pattern can arise if the neutrino mass matrix has the following form,
\begin{equation}
M_{\nu} = \left(
\begin{array}{ccc}
A & B & B \\
B & C & D \\
B & D & C
\end{array}\right) \; .
\end{equation}
This matrix predicts $\sin^{2} 2 \theta_{23} = 1$ and $\theta_{13}=0$, while leaving the value for $\theta_{12}$ undetermined. This mass matrix can arise from an 
underlying $S_{3}$~\cite{Mohapatra:2006pu}, $D_{4}$~\cite{Grimus:2003kq}, or $\mu-\tau$ symmetry~\cite{Fukuyama:1997ky}. A prediction for $\tan^{2}\theta_{12} = 1/2$ arises if  the parameters are chosen such that $A+B = C+D$ is satisfied. 

It has been pointed out that the tri-bimaximal mixing matrix can arise from a family symmetry in the lepton sector based on $A_{4}$~\cite{Ma:2001dn}, which automatically gives rise to, $A+B = C+D$, leading to a prediction for the solar mixing angle, $\sin^{2}\theta_{12} = 1/3$. However, due to its lack of doublet representations, CKM matrix is an identity in most $A_{4}$ models. It is hence not easy to implement $A_{4}$ as a family symmetry for both quarks and leptons~\cite{Ma:2006wm}.  

\subsection{A Realistic SUSY $SU(5) \times T^{\prime}$ Model}

In \cite{Chen:2007afa,Chen:2009gf}, a grand unified model based on SU(5) combined with the double tetrahedral group~\cite{Chen:2007afa,Frampton:1994rk}, $T^{\prime}$, was constructed, which successfully gives rise to near tri-bimaximal leptonic mixing as well as realistic CKM matrix elements for the quarks. The group $T^{\prime}$ is the double covering group of $A_{4}$. In addition to the $1, \; 1^{\prime}, \; 1^{\prime\prime}$ and $3$ representations that $A_{4}$ has, the group $T^{\prime}$ also has three in-equivalent doublet representations, $2, \; 2^{\prime}, \; 2^{\prime\prime}$. This enables the $(1+2)$ assignments, which has been shown to give realistic masses and mixing pattern in the quark sector~\cite{so10ref}. 

One special property of $T^{\prime}$ is the fact that its Clebsch-Gordan coefficients are intrinsically complex, independent of the basis for the two group generators. This thus affords the possibility that CP violation can be entirely geometrical in origin~\cite{Chen:2009gf}.

The charge assignments of various fields in our model are summarized in Table~\ref{tbl:charge},  
\begin{table}
\begin{tabular}{|c|ccc|ccc|} \hline
 & $T_{3}$ & $T_{a}$ & $\overline{F}$ & $H_{5}$ & $H_{\overline{5}}^{\prime}$ & $\Delta_{45}$ 
   \\ \hline
SU(5) & 10 & 10 & $\overline{5}$ & 5 &  $\overline{5}$ & 45 
  \\ \hline
$T^{\prime}$ & 1 & $2$ & 3 & 1 & 1 & $1^{\prime}$ 
\\ \hline
$Z_{12}$ & $\omega^{5}$ & $\omega^{2}$ & $\omega^{5}$ & $\omega^{2}$ & $\omega^{2}$ & $\omega^{5}$ 
 \\ \hline
$Z_{12}^{\prime}$ & $\omega$ & $\omega^{4}$ & $\omega^{8}$ & $\omega^{10}$ & $\omega^{10}$ & $\omega^{3}$ 
\\  \hline
\end{tabular}
\begin{tabular}{|c|cccccc|cc|} \hline
& $\phi$ & $\phi^{\prime}$ & $\psi$ & $\psi^{\prime}$ & $\zeta$ & $N$ & $\xi$ & $\eta$  
\\ 
\hline
SU(5) & 1 & 1 & 1 & 1& 1 & 1 & 1 & 1
\\ 
\hline
$T^{\prime}$ & 3 & 3 & $2^{\prime}$ & $2$ & $1^{\prime\prime}$ & $1^{\prime}$ & 3 & 1 
\\ 
\hline 
$Z_{12}$ & $\omega^{3}$ & $\omega^{2}$ & $\omega^{6}$ & $\omega^{9}$ & $\omega^{9}$ 
& $\omega^{3}$ & $\omega^{10}$ & $\omega^{10}$ 
\\ 
\hline
$Z_{12}^{\prime}$ &  $\omega^{3}$ & $\omega^{6}$ & $\omega^{7}$ & $\omega^{8}$ & $\omega^{2}$ & $\omega^{11}$ & 1 & $1$ 
\\ \hline   
\end{tabular}
\vspace{-0.in}
\caption{Field content of our model. The $Z_{12}$ charges are given in terms of the parameter $\omega = e^{i\pi/6}$.}  
\label{tbl:charge}
\end{table}
Due to the transformation properties of various fields, only top quark mass is allowed by the $T^{\prime}$ symmetry, and thus it is the only mass term that can be generated at the renormalizable level. To give masses to the lighter generations of fermions, which transform non-trivially under $T^{\prime}$, the $T^{\prime}$ symmetry has to be broken, which is achieved by a set of flavon fields. Due to the presence of the  $Z_{12} \times Z_{12}^{\prime}$ symmetry, only ten operators are allowed in the model, and hence the model is very predictive, the total number of parameters being nine in the Yukawa sector for the charged fermions and the neutrinos. The Lagrangian of the model is given as follows,
\begin{eqnarray}
\mathcal{W}_{\mbox{\tiny Yuk}} \!\! & \!\! = \!\! & \!\!  \mathcal{W}_{TT} + \mathcal{W}_{TF} + \mathcal{W}_{\nu} \; , \\
\mathcal{W}_{TT} \!\!  & \!\! = \!\! & \!\! y_{t} H_{5} T_{3} T_{3} + \frac{1}{\Lambda^{2}}  H_{5} \biggl[ y_{ts} T_{3} T_{a} \psi \zeta 
\nonumber \\
& & + y_{c} T_{a} T_{b} \phi^{2} \biggr] 
+ \frac{1}{\Lambda^{3}} y_{u} H_{5} T_{a} T_{b} \phi^{\prime 3}\; , \label{eq:Ltt} \\ 
\mathcal{W}_{TF} \!\! & \!\! = \!\! & \!\!  \frac{1}{\Lambda^{2}} y_{b} H_{\overline{5}}^{\prime} \overline{F} T_{3} \phi \zeta + \frac{1}{\Lambda^{3}} \biggl[ y_{s} \Delta_{45} \overline{F} T_{a} \phi \psi \zeta^{\prime}  \nonumber \\
& & \quad + y_{d} H_{\overline{5}^{\prime}} \overline{F} T_{a} \phi^{2} \psi^{\prime} \biggr]  \; ,  \label{eq:Ltf} \\
\mathcal{W}_{\nu} \!\! & \!\! = \!\! & \!\!  \lambda_{1} NNS \nonumber \\
& & +  \frac{1}{\Lambda^{3}} \biggl[ H_{5}  \overline{F} N \zeta \zeta^{\prime} \biggl( \lambda_{2} \xi 
+ \lambda_{3} \eta\biggr) \biggr] \; ,
\label{eq:Lff}
 \end{eqnarray}
 where $\Lambda$ is the $T^{\prime}$ symmetry breaking scale.  (For the VEV's of various scalar fields, see Ref.~\cite{Chen:2007afa}.) The parameters $y$'s and $\lambda$'s are the coupling constants.

The interactions in $\mathcal{W}_{TT}$ and $\mathcal{W}_{TF}$ gives rise to the up-type quark and down-type quark  mass matrices, $M_{u}$ and $M_{d}$, respectively. Since the lepton doublets and iso-singlet down-type quarks are unified into a $\overline{5}$ of $SU(5)$, their mass matrices are related. Upon the breaking of $T^{\prime}$ and the electroweak symmetry, these mass matrices are given in terms of seven parameters by~\cite{Chen:2007afa},
\begin{eqnarray}
\! \!  \! \!  \! \!  \! \!  \! \!  \! \!   \frac{M_{u}}{y_{t} v_{u}} \! \! & \! \! = \! \! & \! \! \left( \begin{array}{ccccc}
i g & ~~ &  \frac{1-i}{2}  g & ~~ & 0\\
\frac{1-i}{2} g & & g + (1-\frac{i}{2}) h  & & k\\
0 & & k & & 1
\end{array}\right)  , \\
\! \!  \! \!  \! \!  \! \!  \! \!  \! \!   \frac{M_{d}, \; M_{e}^{T}}{y_{b} v_{d} \phi_{0}\zeta_{0}} \! \!  & \! \!  = \! \! & \! \!  \left( \begin{array}{ccccc}
0 & ~~ & (1+i) b & ~~ & 0\\
-(1-i) b & & (1,-3) c & & 0\\
b & &b & & 1
\end{array}\right)  \; , 
\end{eqnarray}
where $a, \; b, ...$, {\it etc}, are given in terms of the flavon VEVs. 
The $SU(5)$ relation, $M_{d} = M_{e}^{T}$, is manifest,  
except for the factor of $-3$ in the (22) entry of $M_{e}$, due to the $SU(5)$ CG coefficient through the coupling to $\Delta_{45}$. In addition to this $-3$ factor, the Georgi-Jarlskog (GJ) relations at the GUT scale,
$m_{e} \simeq  \frac{1}{3} m_{d}$, $m_{\mu}  \simeq 3 m_{s}$, and  
$m_{\tau} \simeq  m_{b}$, 
also require $M_{e,d}$ being non-diagonal, leading to corrections to the TBM pattern~\cite{Chen:2007afa}.  Note that the complex coefficients in the above mass matrices arise {\it entirely} from the CG coefficients of the $T^{\prime}$ group theory. More precisely, these complex CG coefficients appear in couplings that involve doublet representations of $T^{\prime}$.

The complex mass matrices $M_{u,d}$ lead to a complex quark mixing matrix, $V_{CKM} = V_{u,L}^{\dagger} V_{d,L}$. 
The relation $\theta_{c} \simeq | \sqrt{m_{d}/m_{s}} - \sqrt{m_{u}/m_{c}}|  \simeq \sqrt{m_{d}/m_{s}}$, 
  is manifest in our model.
 Similarly, the mixing angle $\theta_{12}^{e}$ in the diagonalization matrix $V_{e,L}$ for the charged lepton sector is given by,
$\theta_{12}^{e} \simeq  \sqrt{m_{e}/m_{\mu}}$. 
 Using the Georgi-Jarlskog relations, one then obtains the following relation between the Cabibbo angle and the mixing angle $\theta_{12}^{e}$ in the charged lepton sector,
 $\theta_{12}^{e} \simeq \frac{1}{3} \theta_{c}$.
All other elements in $V_{e,L}$ are higher order in $\theta_{c}$, and hence $\theta_{12}^{e}$ gives the dominant corrections to the TBM mixing pattern.  

Due to the discrete symmetries in our model,  the mass hierarchy arises dynamically without invoking an additional U(1) symmetry. The $Z_{12}$ symmetry also forbids Higgsino-mediated proton decays in SUSY version of the model. Due to the $T^{\prime}$ transformation property of the matter fields, the $b$-quark mass can be generated only when the $T^{\prime}$ symmetry is broken, which naturally explains  the hierarchy between $m_{b}$ and $m_{t}$. The $Z_{12} \times Z_{12}^{\prime}$ symmetry, to a very high order, also forbids operators that lead to nucleon decays. 

The interactions in $\mathcal{W}_{\nu}$ lead to the following neutrino mass matrix, 
\begin{eqnarray}\label{eq:fd} \! \! \! \! \! \! \! \! \! \! \! \! \! 
\frac{M_{RR}}{S_{0}} \!\! & \!\! = \!\! & \!\!  \left( \begin{array}{ccc}
1 & 0 & 0 \\
0 & 0 & 1 \\
0 & 1 & 0 
\end{array}\right)  \; , \\ 
\frac{M_{D}}{ \zeta_{0} \zeta^{\prime}_{0} v_{u}} \!\! & \!\! = \!\! & \!\! \left( \begin{array}{ccc}
2\xi_{0} + \eta_{0} & -\xi_{0} & -\xi_{0} \\
-\xi_{0} & 2\xi_{0} & -\xi_{0} + \eta_{0} \\
-\xi_{0} & -\xi_{0}+\eta_{0} & 2\xi_{0} 
\end{array}\right) ,
\end{eqnarray}
which is parametrized by {\it two} parameters, giving the three absolute neutrino masses~\cite{Chen:2007afa}. As these interactions involve only the triplet representations of $T^{\prime}$, all CG coefficients are real, leading to a real neutrino Majorana mass matrix. The effective neutrino mass matrix,  $M_{\nu} = M_{D} M_{RR}^{-1} M_{D}^{T}$, has the special property that it is form diagonalizable, {\it i.e.} independent of the values of $\xi_{0}$ and $u_{0}$, it is diagonalized by the tri-bimaximal mixing matrix,
\begin{eqnarray}
\! \! \! \! \! \! \! \! \! \! \! \! \! 
U_{\mbox{\tiny TBM}}^{T} M_{\nu} U_{\mbox{\tiny TBM}} \! \! & \! \! \equiv \! \! & \! \!    \mbox{diag} (m_{1}, m_{2}, m_{3})   \; , 
\\
\! \! \! \! \! & \! \! = \! \! & \! \!  \mbox{diag}((u_{0} + 3 \xi_{0})^{2}, u_{0}^{2}, -(-u_{0}+3\xi_{0})^{2}) \frac{v^{\prime 2}}{S_{0}} 
\nonumber \; ,
\end{eqnarray}
where $v^{\prime 2} = \zeta_{0}\zeta_{0}^{\prime} v_{u}$. (For general conditions for form diagonalizability, see Ref.~\cite{Chen:2009um}.) 
While the neutrino mass matrix is real, the complex charged lepton mass matrix $M_{e}$, leads to a complex 
$V_{\mbox{\tiny PMNS}} = V_{e, L}^{\dagger} U_{\mbox{\tiny TBM}}$. 
The Georgi-Jarlskog relations for three generations are obtained. This inevitably requires non-vanishing mixing in the charged lepton sector, as mentioned previously, leading to corrections to the tri-bimaximal mixing pattern.  Consequently, our model predicts a non-vanishing $\theta_{13}$, which is related to the Cabibbo angle as, 
\begin{equation}
\theta_{13}\sim \theta_{c}/3\sqrt{2} \; . 
\end{equation}  
Numerically, this is close to $\sin\theta_{13} \sim 0.05$ which is accessible to the Daya Bay reactor experiment. In addition, our model gives rise to a sum rule between the Cabibbo and the solar mixing angle for the neutrinos, 
\begin{equation}
\tan^{2}\theta_{\odot} \simeq \tan^{2} \theta_{\odot, \mathrm{TBM}} + \frac{1}{2} \theta_{c} \cos\delta_{\ell} \; ,
\end{equation} 
which is a consequence of the Georgi-Jarlskog relations in the quark sector. Here the parameter $\delta_{\ell}$ is the Dirac CP phase in the lepton sector in the standard parametrization. This deviation could account for the difference between the experimental best fit value for the solar mixing angle and the value predicted by the tri-bimaximal mixing matrix. 

Since the three absolute neutrino mass eigenvalues are determined by only two parameters, there is a sum rule that relates the three light masses, 
\begin{equation}
\left| |\sqrt{m_{1}}|  + |\sqrt{m_{3}}| \right| =  2 |\sqrt{m_{2}}| \; ,
\end{equation}
for $(3\xi_{0} + \eta_{0})(3\xi_{0} - \eta_{0}) > 0$, and 
\begin{equation}
\left| |\sqrt{m_{1}}|  - |\sqrt{m_{3}}| \right| =  2 |\sqrt{m_{2}}| \; ,
\end{equation}
for $(3\xi_{0} + \eta_{0})(3\xi_{0} - \eta_{0}) < 0$.  
In the model, both normal and inverted hierarchical mass patterns can be accommodated. This is to be contrasted with the case using the effective operator of the $HHLL$ type, in which the normal hierarchy is predicted~\cite{Chen:2007afa}. 

In our model, all 22 observable fermion masses and mixing angles and CP violating measures are fitted with only 9 parameters.

\subsection{Numerical Results}

With the input parameters
$b \equiv \phi_{0} \psi^{\prime}_{0} /\zeta_{0} = 0.00304$, $c\equiv \psi_{0}N_{0}/\zeta_{0}=-0.0172$, 
$k \equiv y^{\prime}\psi_{0}\zeta_{0}=-0.0266$,  
$h\equiv \phi_{0}^{2}=0.00426$, and $g \equiv \phi_{0}^{\prime 3}= 1.45\times 10^{-5}$, 
the following mass ratios are obtained,
$m_{d}: m_{s} : m_{b} \simeq \theta_{c}^{\scriptscriptstyle 4.7} : \theta_{c}^{\scriptscriptstyle 2.7} : 1$ and  
$m_{u} : m_{c} : m_{t} \simeq \theta_{c}^{\scriptscriptstyle 7.5} : \theta_{c}^{\scriptscriptstyle 3.7} : 1$, 
with $\theta_{c} \simeq \sqrt{m_{d}/m_{s}} \simeq 0.225$. We have also taken $y_{t} = 1.25$ and $y_{b}\phi_{0} \zeta_{0} \simeq m_{b}/m_{t} \simeq 0.011$ and have taken into account the renormalization group corrections. As a result  of the Georgi-Jarlskog relations, realistic charged lepton masses are obtained. 
These parameters also gives rise to the following complex CKM matrix,
\begin{eqnarray}\!\!\!\!\!\!
\tiny{
\left( \begin{array}{ccc}
0.974 \!\!\! & \!\!\! 0.227  & 0.00412e^{-i45.6^{o}} \\
-0.227 - 0.000164 e^{i45.6^{o}} & 0.974 - 0.0000384 e^{i45.6^{o}} & 0.0411 \\
0.00932 - 0.00401 e^{i45.6^{o}} & -0.0400 - 0.000935 e^{i45.6^{o}} & 1
\end{array}\right).} 
\end{eqnarray}
The predictions of our model for the angles in the unitarity triangle  and the Jarlskog invariant in the quark sector are,
$\beta = 23.6^{o}$, 
$\alpha = 110^{o}$, 
$\gamma = \delta_{q} = 45.6^{o}$, and 
$J = 2.69 \times 10^{-5}$, 
where $\delta_{q}$ is the CP phase in the standard parametrization, which has a large experimental uncertainty at present. 
In terms of the Wolfenstein parameters, we have 
$\lambda = 0.227$, $A = 0.798$, 
$\overline{\rho} = 0.299$, and 
$\overline{\eta} = 0.306$.
These predictions are consistent with the current experimental limits at $3\sigma$. 
 
In the lepton sector,  the diagonalization matrix for the charged lepton mass matrix combined with $U_{TBM}$ gives numerically the following PMNS matrix, 
\begin{equation}\!\!\!\!\!\!
\tiny{
\left( \begin{array}{ccc}
0.838  & 0.542 & 0.0583 e^{-i227^{o}}   \\
-0.385  - 0.0345 e^{i227^{o}} & 0.594 - 0.0224 e^{i227^{o}} & 0.705   \\
0.384 - 0.0346 e^{i227^{o}} & -0.592 - 0.0224 e^{i227^{o}} & 0.707
\end{array}\right) \; ,}
\end{equation}
which predicts 
$\sin^{2}\theta_{\mathrm{\scriptscriptstyle atm}} = 1$,  
$\tan^{2}\theta_{\odot} = 0.420$, and 
$|U_{e3}| = 0.0583$. 
The two VEV's, $\eta_{0} = 0.1707$, $\xi_{0} = -0.0791$, and $S_{0} = 10^{12}$ GeV, give 
$ \Delta m_{atm}^{2} = 2.5 \times 10^{-3} \; \mbox{eV}^{2}$ and 
 $\Delta m_{\odot}^{2} = 7.6 \times 10^{-5} \; \mbox{eV}^{2}$. 
 The leptonic Jarlskog is predicted to be $J_{\ell} = -0.00967$, 
 and equivalently, this gives  a Dirac CP phase, 
 \begin{equation}
 \delta_{\ell} = 227^{o} \; , 
 \end{equation}
 which is very close to the current best fit value of $\delta_{\ell} = 220^{o}$ from SuperK~\cite{SuperK-Nu2010}. 
 With such $\delta_{\ell}$, the correction from the charged lepton sector can account for the difference between the TBM prediction and  the current best fit value for $\theta_{\odot}$. Our model predicts 
\begin{equation}\!\!\!\!\!\!\!\!
(m_{1}, \; m_{2}, \; m_{3}) = (0.00134, \; 0.00882, \; 0.0504) \;  \mbox{eV} \; , \; 
\end{equation}
for normal hierarchy, with Majorana phases 
\begin{eqnarray}
\alpha_{21} = 0 \; , \quad \alpha_{31} = \pi \; . 
\end{eqnarray}

Since the leptonic Dirac CP phase, $\delta_{\ell}$, is the only non-vanishing CP violating phase in the lepton sector, a connection~\cite{Chen:2007fv,Chen:2004ww} between leptogenesis and low energy CP violating leptonic processes, such as neutrino oscillation, can exist in our model.

\subsection{Curing the FCNC Problem: Family Symmetry v.s. Minimal Flavor Violation}

We note that in addition to the capability of giving rise to mixing angles and CP violation from CG coefficients, the group $T^{\prime}$ has recently been utilized in a Randall-Sundrum  model to avoid tree-level flavor-changing neutral currents~\cite{Chen:2009gy}, which are present in generic RS models.

\section{Beyond Tri-bimaximal Mixing}

With the current experimental precision on the measurements of the neutrino mixing angles, the TBM mixing pattern can be accidental~\cite{Abbas:2010jw}. Other mixing patterns~\cite{Albright:2010ap} that haven been suggested are the following.

\subsection{Golden Ratio for Solar Mixing Angle}

It has been suggested~\cite{Datta:2003qg} that the solar mixing angle is related to the Golden ratio,
\begin{equation} \!\!\!\!\!\!
\tan^{2}\theta_{\odot} = 1/ \Phi^{2} = 0.382, \;
\Phi = (1 + \sqrt{5})/2 = 1.62 \; , \;
\end{equation}
which is $1.4\sigma$ below the experimental best fit value. Symmetries based on $Z_{2}\times Z_{2}$~\cite{Kajiyama:2007gx}, $A_{5}$~\cite{Kajiyama:2007gx,Everett:2008et}, and $D_{10}$~\cite{Adulpravitchai:2009bg}, have been utilized to get the above relation.

\subsection{Dodeca Neutrino Mixing Matrix}

It has also been suggested that~\cite{Kim:2010zu} that to the leading order $\theta_{c} = 15^{o}$, $\theta_{\odot} = 30^{o}$, $\theta_{\mathrm{\scriptscriptstyle atm}} = 45^{o}$. This corresponds to the following PMNS matrix,
\begin{equation}
V_{PMNS} = \left(\begin{array}{ccc}
\cos\frac{\pi}{6} & \sin\frac{\pi}{6} & 0 \\
-\frac{1}{\sqrt{2}}\sin\frac{\pi}{6} & \frac{1}{\sqrt{2}} \cos\frac{\pi}{6} & -\frac{1}{\sqrt{2}} \\
-\frac{1}{\sqrt{2}}\sin\frac{\pi}{6} & \frac{1}{\sqrt{2}} \cos\frac{\pi}{6} & -\frac{1}{\sqrt{2}} \end{array}
\right) \; ,
\end{equation}
and can arise from the $Z_{2} \times Z_{12} = D_{12}$ symmetry. Corrections to the predictions for the mixing angles can arise from $D_{10}$ symmetry breaking.

\section{TeV Scale Seesaw Models and $U(1)^{\prime}$ Symmetry}

In the conventional wisdom, the smallness of the neutrino masses is tied to the high scale of the new physics that generates neutrino masses as described in the above scenarios.  In \cite{Chen:2006hn}, an alternative was proposed in which the small neutrino masses are generated with TeV scale physics. This allows the possibility of testing the new physics that gives rise to neutrino masses at the Tevatron and the LHC. This is achieved by augmenting the Standard Model with a non-anomalous $U(1)^{\prime}$ symmetry and 3 right-handed neutrinos. Due to the presence of  the $U(1)^{\prime}$ symmetry, neutrino masses can only be generated by operators with very high dimensionality, which in turn allows a seesaw scale as low as a TeV. By measuring the decay properties of the $Z^{\prime}$ gauge boson, the model can be tested at the LHC~\cite{Chen:2009fx}. 

While there exists an earlier claim~\cite{Ibanez:1994ig} that the U(1) symmetry has to be anomalous in order to generate realistic fermion masses and mixing, we note that counter examples to this claim have been found~\cite{Chen:2008tc} in which it is shown that a non-anomalous $U(1)^{\prime}$  symmetry can be a family symmetry giving rise to realistic masses and mixing angles of the SM fermions. In addition to solving the fermion mass problem, such a non-anomalous $U(1)^{\prime}$ family symmetry also provides a solution to the tachyonic slepton mass problem which generically is present in anomaly mediated SUSY breaking~\cite{Chen:2010tf}.

\section{Conclusion}

We present a model based on $SU(5)$ and the double tetrahedral group $T^{\prime}$ as the family symmetry. CP violation in our model is entirely geometrical due to the presence of the complex group theoretical CG coefficients in $T^{\prime}$. The Georgi-Jarlskog relations automatically lead to a sum rule between the Cabibbo angle and the solar mixing angle for the neutrino. The predicted CP violation measures in the quark sector are consistent with the current experimental data. The leptonic Dirac CP violating phase is predicted ~\cite{Chen:2009gf} to be $\delta_{\ell} \sim 227^{o}$, which is very close to the current best fit value of $\delta_{\ell} = 220^{o}$ from SuperK and it gives the cosmological matter-antimatter asymmetry~\cite{Chen:2009}.

\section*{Acknowledgments}
The work of M-CC was supported, in part, by the National Science Foundation under Grant No. PHY-0709742 and PHY-0970173. 
The work of KTM was supported, in part, by the Department of Energy under Grant No. DE-FG02-04ER41290.





\bibliographystyle{elsarticle-num}
\bibliography{}



\end{document}